%% file: manuscript_arxiv.tex
\documentclass[english]{elsarticle}

\usepackage{hyperref}

\usepackage[T1]{fontenc}
\usepackage[latin9]{inputenc}
\usepackage{geometry}
\usepackage{graphicx}
\usepackage{fancyhdr}
\makeatletter
\usepackage{graphicx}
\providecommand{\tabularnewline}{\\}
\usepackage{booktabs}

\makeatother

\usepackage{babel}
\begin{document}
\begin{frontmatter}

\title{Exploring RNA structure and dynamics through enhanced sampling simulations}

\author{Vojt\v{e}ch Ml\'{y}nsk\'{y}}
\ead{vojtech.mlynsky@sissa.it}
\author{Giovanni Bussi\corref{cor1}}
\ead{bussi@sissa.it}
\cortext[cor1]{Corresponding author}

\address{Scuola Internazionale Superiore di Studi Avanzati, SISSA,}

\address{via Bonomea 265, 34136 Trieste, Italy}
\begin{abstract}
RNA function is intimately related to its structural dynamics. Molecular
dynamics simulations are useful for exploring biomolecular flexibility
but are severely limited by the accessible timescale. Enhanced sampling
methods allow this timescale to be effectively extended in order to
probe biologically-relevant conformational changes and chemical reactions.
Here, we review the role of enhanced sampling techniques in the study of RNA systems.
We discuss the challenges and promises associated with the application of these methods
to force-field validation, exploration of conformational landscapes and ion/ligand-RNA interactions,
as well as catalytic pathways.
Important technical aspects of these methods, such
as the choice of the biased collective variables and the analysis
of multi-replica simulations, are examined in detail. Finally, a
perspective on the role of these methods in the characterization of
RNA dynamics is provided.
\end{abstract}
\end{frontmatter}

\vspace{\fill}

\copyright~2017. This manuscript version is made available under the CC-BY-NC-ND 4.0 license http://creativecommons.org/licenses/by-nc-nd/4.0/

\newpage

\section*{Introduction}

Ribonucleic acids (RNA) play fundamental roles in the cell, ranging
from catalysis \cite{lilley2017rna} to control of gene expression
\cite{morris2014riseregRNA}. RNA function is often linked to its
three-dimensional structure, typically obtained using X-ray crystallography
\cite{westhof2015twenty} or nuclear magnetic resonance (NMR) \cite{rinnenthal2011mapping}.
However, RNA molecules are not static and might exhibit a multitude
of accessible functional structures in a narrow energetic range. Many
examples have been reported, ranging from flexible RNA motifs \cite{lilley2012structure}
to excited states \cite{dethoff2012visualizing} and, in the extreme
case, riboswitches \cite{serganov2013decade}. In addition, RNA catalysis
is initiated by a pre-organization of the active site, and transition
states (TSs) need to be stabilized by neighboring groups \cite{emilsson2003ribozyme}.
The mentioned events might require timescales ranging from microseconds
to seconds or hours to be observed in an experimental setup.

Molecular dynamics (MD) simulations, both using empirical force fields
and quantum mechanics/molecular mechanics approaches (QM/MM), are
in principle a powerful tool to access RNA flexibility. However, they
are limited to timescales of a few microseconds (for empirical force
fields) or a few hundreds of picoseconds (for QM/MM-MD approaches).
In order to address the conformational transitions and chemical reactions
mentioned above, they should be complemented with enhanced sampling
methods. Even dedicated machines capable to perform millisecond-long classical MD
need enhanced sampling methods in order
to access biologically relevant timescales \cite{pan2016demonstrating}.

We here present a survey on the recent applications of enhanced sampling
techniques to atomistic MD simulations of RNA systems. Many recent
reviews discuss in detail enhanced sampling methods \cite{bernardi2015enhanced,valsson2016enhancing}
and MD simulations of RNA \cite{vsponer2013understand,huang2015nucleic,vsponer2017understand,vangaveti2017advances,smith2017physics}.
We opted for proceeding in an orthogonal direction, highlighting which
enhanced sampling methods have been recently applied to RNA systems
and, at the same time, underlying which aspects of RNA dynamics can
benefit of enhanced sampling methods. In order to take a picture of
the current state of the art for the application of these techniques
to RNA systems, we deliberately limited the survey to the past two
years. In addition, we only considered atomistic explicit solvent simulations where hydrogen atoms
and water molecules are explicitly included.

\paragraph*{Basic Assumptions}

A fundamental issue in MD simulations is the choice of an appropriate
model to compute the interatomic forces. This is done using empirical
force fields (see \cite{vsponer2017understand,vangaveti2017advances,smith2017physics}
and references therein) and/or QM methods. In the latter case, a compromise
between accuracy and computational cost should be found, choosing
between fast but approximate semi-empirical (SE) methods and
more accurate but computationally demanding
density functional theory (DFT) methods (see \cite{vsponer2013understand,huang2015nucleic}
and references therein).

\paragraph*{Enhanced Sampling}

\begin{figure}
\begin{centering}
\includegraphics[width=0.7\textwidth]{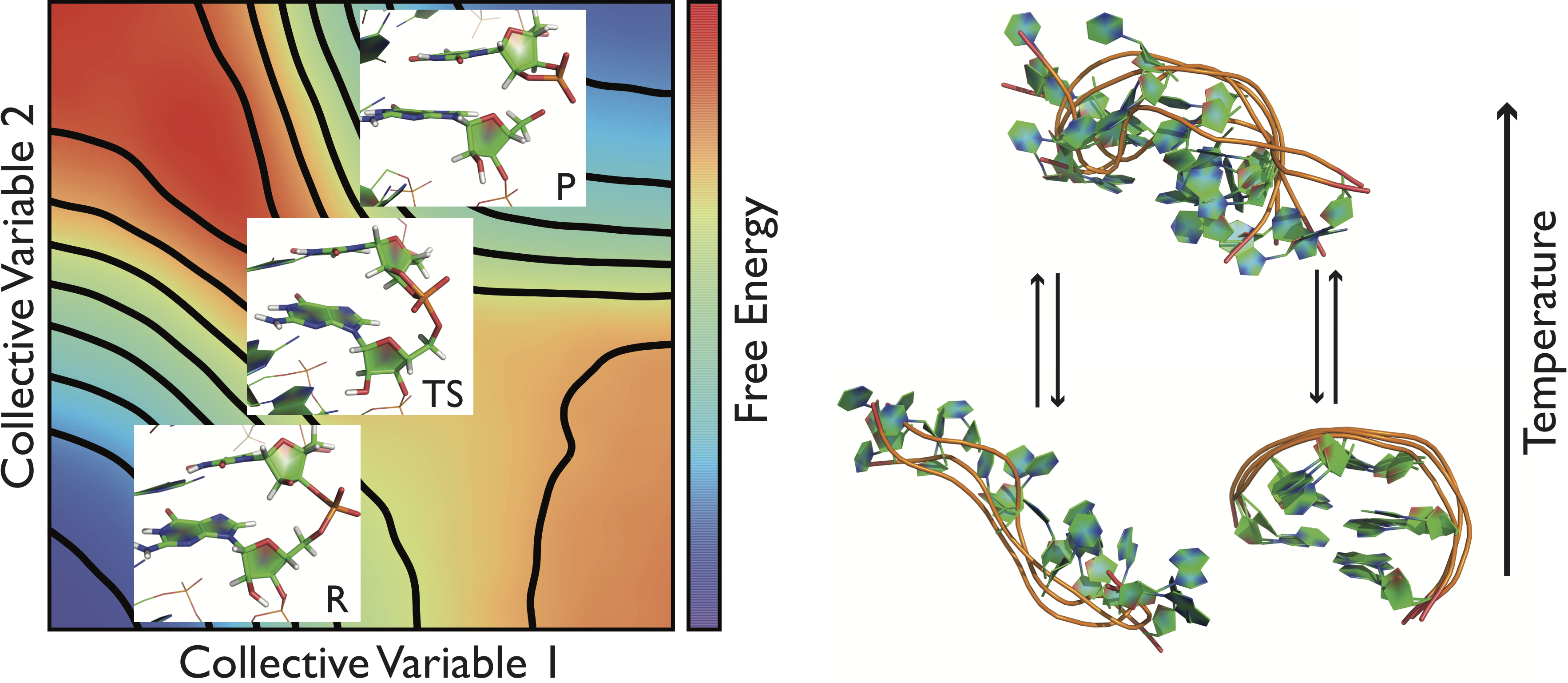}
\par\end{centering}
\caption{Scheme representing distinctive features of enhanced sampling methods of different classes.
Methods based on collective variables (CVs), where a small number
of CVs capable to describe the important free-energy minima (\emph{e.g.},
reactant (R), transition (TS), and product (P) states), are identified
\emph{a priori} (panel on the left). These variables are then exploited
to enhance sampling. Methods based on tempering, where the temperature
of the system is repeatedly increased and decreased (panel on the
right). The increased conformational dynamics at high temperature
allows more conformations to be explored also at low temperature.\label{fig:Schemes-representing-enhanced}}
\end{figure}

A central idea of all enhanced sampling methods is to alter the system's
dynamics to characterize specific events that would otherwise require
significantly longer simulation timescales. Generally speaking, this
can be done in two ways (Figure \ref{fig:Schemes-representing-enhanced}):
(i) by changing the probability distribution of a limited number of
selected degrees of freedom, so called collective variables (CVs),
deemed to be important for the investigated conformational transition;
(ii) by acting on the total energy or, equivalently, on the temperature
of the system. Prototypical methods for these two approaches are (i)
umbrella sampling (US) and (ii) temperature replica-exchange molecular
dynamics (T-REMD), respectively (see \cite{bernardi2015enhanced,valsson2016enhancing}
and references therein). Methods based on CVs are extremely efficient
when the chosen CVs identify correctly the kinetically relevant states
of the system, including metastable and TSs. Methods based on tempering
are more computationally demanding, but usually require less \emph{a priori }information.
CV-based and tempering methods can be combined and methods at the
boundary between these two classes have been proposed as well. We
note that the usage of replicas is not necessarily a distinctive trait
of tempering methods. US can indeed be performed in a replica-exchange
scheme, as it is discussed below. Conversely, temperature methods
using a single simulation are used as well. Alchemical approaches
such as the free-energy perturbation method, where transitions are
enforced through a non-physical path involving changes in particle
number and/or identity \cite{bernardi2015enhanced}, can be considered
as a special case of CV-based methods. Other approaches using unbiased
simulations to analyze and reconstruct long-time kinetics, as well
as non-dynamical methods where energies of individual structures are
calculated and compared, are not considered here.

\section*{Applications of Enhanced Sampling Methods to RNA Systems}

\begin{table*}
\resizebox{\textwidth}{!}{

\input{tableshort.tex}

}

\caption{Summary of recent application of enhanced sampling methods to RNA
systems. Explanation of the acronyms and additional information are
included in Table S1 in Supporting Information.\label{tab:Summary-of-recent}
The online version of this table will be kept up to date at \url{https://github.com/srnas/rna-enhanced-sampling}.
}
\end{table*}

Table \ref{tab:Summary-of-recent} reports an extensive list of
publications in the last two years where enhanced sampling methods were applied to RNA systems.
We arbitrarily classified them in groups according to the presented
application, although some of them could be assigned to more than
one group.

\paragraph*{Refinements and Validations of Force Fields}

Historically, force fields have been validated by analyzing plain
MD simulations starting from the native structure. Taking advantage
of enhanced sampling techniques, small RNA motifs can be sampled until
statistical convergence is reached in order to expose all the potential force field limitations. In several studies listed
in this category, the population of the native structure (or the relative population of a number of structures) was
computed and compared with experiments \cite{bergonzo2015highly,bergonzo2015improved,Kuhrova2016,bottaro2016free,yang2016predicting,aytenfisu2017revised}.
The direct comparison with raw experimental data is more suitable for small
unstructured RNAs \cite{gil2016empirical,bottaro2016free,cesari2016maxent}.
Ref. \cite{sakuraba2015predicting} encouragingly predicted the effect
of simple mutations on the dimerization energy of a duplex in reasonable
agreement with experimental data. Refs. \cite{gil2016empirical,cesari2016maxent} used enhanced
sampling methods during the construction of the force field rather than just to validate it.
Some works of this category \cite{bergonzo2015highly,Kuhrova2016,bottaro2016free}
suggest that none of the available
force fields yet allows predictive folding of RNA hairpin loops or
larger systems to be performed reliably.
One may wonder that the requirement to obtain a stable folded structure with all native hydrogen
bonds simultaneously formed might be too restrictive for some motifs. However,
a direct comparison with NMR data in ref. \cite{bottaro2016free} emphasized that, at least for one of the investigated tetraloops,
using a too loose criterion would lead to structures that are not compatible with solution experiments being reported as correct.

\paragraph*{Conformational Landscapes}

This is the wider group considered and includes papers discussing
the conformational landscape of systems ranging from individual nucleosides
\cite{radak2015characterization,larsen2015thermodynamic,yildirim2015computational},
small loops \cite{gil2015,haldar2015insights,miner2016free,miner2017equilibrium},
duplexes \cite{park2015crystallographic,verona2017focus,pan2017structure},
stem-loops \cite{bergonzo2015stem,yildirim2015computational,wu2015multivalent,takahashi2016using,borkar2016structure,borkar2017simultaneous,pathak2017water,sun2017protonation},
and pseudoknots \cite{bian2015free}, up to larger aptamers \cite{hayatshahi2017investigating,warfield2017molecular},
riboswitches \cite{xue2015folding,di2015kissing}, RNA:peptide \cite{borkar2016structure},
and RNA:protein complexes \cite{vukovic2016molecular,palermo2017crispr}.
None of the applications to large RNA systems is designed to sample
extensively the conformational space, which would be extremely expensive
and probably counterproductive considering the force-field limitations discussed
above. However, local excitations can provide a wealth of information
that can be compared with experiment. Interestingly, in some of these
works the simulation is complemented with experimental data in order
to improve the accuracy of the resulting structural ensemble \cite{borkar2016structure,borkar2017simultaneous}.

\paragraph*{Ion Interactions and Ion/ligand Induced Conformational Changes}

Divalent cations are crucial for RNA folding and catalysis. However,
the typical timescale for direct binding of divalent cations on RNA
is on the order of the millisecond, and should thus be simulated using
enhanced sampling methods. Some of the works of this section focus
indeed on interactions between Mg$^{2+}$ ions and individual nucleosides
or RNA structural motifs \cite{panteva2015force,cunha2016ions,lemkul2016characterization,hayatshahi2017computational}.
Other studies address structural reconformations induced by the presence
of a (usually) small rigid molecule (ligand) in a binding pocket and
the related problem of computing the affinity between ligands and
RNA motifs \cite{bochicchio2015molecular,liberman2015structural,hu2017ligand,grasso2017free}.

\paragraph*{Reactivity and Catalysis}

Small self-cleaving ribozymes are interesting model systems for probing
general principles of RNA catalysis. However, the rugged free-energy
landscapes of phosphodiester cleavage reactions present a significant
obstacle in a consistent identification of feasible reaction pathways
in catalytic systems as well as in general `noncatalytic' RNA motifs
\cite{mlynsky2016inline}. Application of enhanced sampling methods
helped to characterize reaction mechanisms in hammerhead \cite{chen2017divalent},
hepatitis delta virus (HDV) \cite{radak2015assessment,thaplyal2015inverse},
twister \cite{gaines2016ribozyme}, group-II intron ribozymes \cite{casalino2016activates}
and glucosamine-6-phosphate synthase (\emph{glmS}) riboswitch \cite{zhang2015role,zhang2016assessing,bingaman2017glcn6p},
where charged nucleobases, Mg$^{2+}$ ions, water molecules and/or
other ligands are involved. Other works reported how changes in external
conditions, \emph{i.e.}, interaction with a mineral surface \cite{swadling2015structure}
or high pressure \cite{schuabb2017pressure}, affect the
pre-organization of the active site required for catalysis.

\paragraph*{General Considerations}

The simulations performed with empirical force fields are typically
sampling timescales ranging from approximately 1 $\mu$s to a few
tens of $\mu$s, which corresponds to a few days up to a few weeks
of computational time considering current hardware and software. Remarkable
efforts have also been reported, such as the extensive benchmark of
force fields performed by Bergonzo \emph{et al.} \cite{bergonzo2015highly,bergonzo2015improved}
and by Kuhrova \emph{et al.} \cite{Kuhrova2016}, which reached or
even surpassed the millisecond timescale of aggregated time. QM methods
are considerably slower and accurate DFT methods are typically used
on the ps time scale. An exception is represented by SE potentials
that can probe the ns timescale. The AMBER force field is by far the
most adopted empirical force field. In some works, alternative modifications
were tested, including modified non-bonded parameters \cite{bergonzo2015highly,bergonzo2015improved,Kuhrova2016,yang2016predicting,miner2016free,takahashi2016using}
or dihedral reparametrizations \cite{cesari2016maxent,aytenfisu2017revised,takahashi2016using,liberman2015structural}.
A limited number of applications used the CHARMM force field, either
for very short simulations \cite{larsen2015thermodynamic} or for
simulations where RNA backbone was constrained \cite{lemkul2016characterization}.
The CHARMM force field has been already reported to lead to unstable
trajectories in plain MD simulations. However, there is a significant
chance to observe spurious states with any of the current force fields
when applying enhanced sampling methods. The quality of the force
field in the MM part of QM/MM simulations is probably less critical
due to the short simulated timescales and to the fact that usually
enhanced sampling methods are employed to accelerate events in the
QM part. DFT functionals are mostly used for QM calculations because
they offer sufficient accuracy for estimating reaction barriers. Some
works benefited from the usage of efficient SE methods that allow
extensive sampling at the expense of some tuning and external corrections
\cite{radak2015characterization,radak2015assessment,mlynsky2016inline}.

\paragraph*{Enhanced Sampling Methods}

\begin{figure}
\centering{}\includegraphics[width=0.7\textwidth]{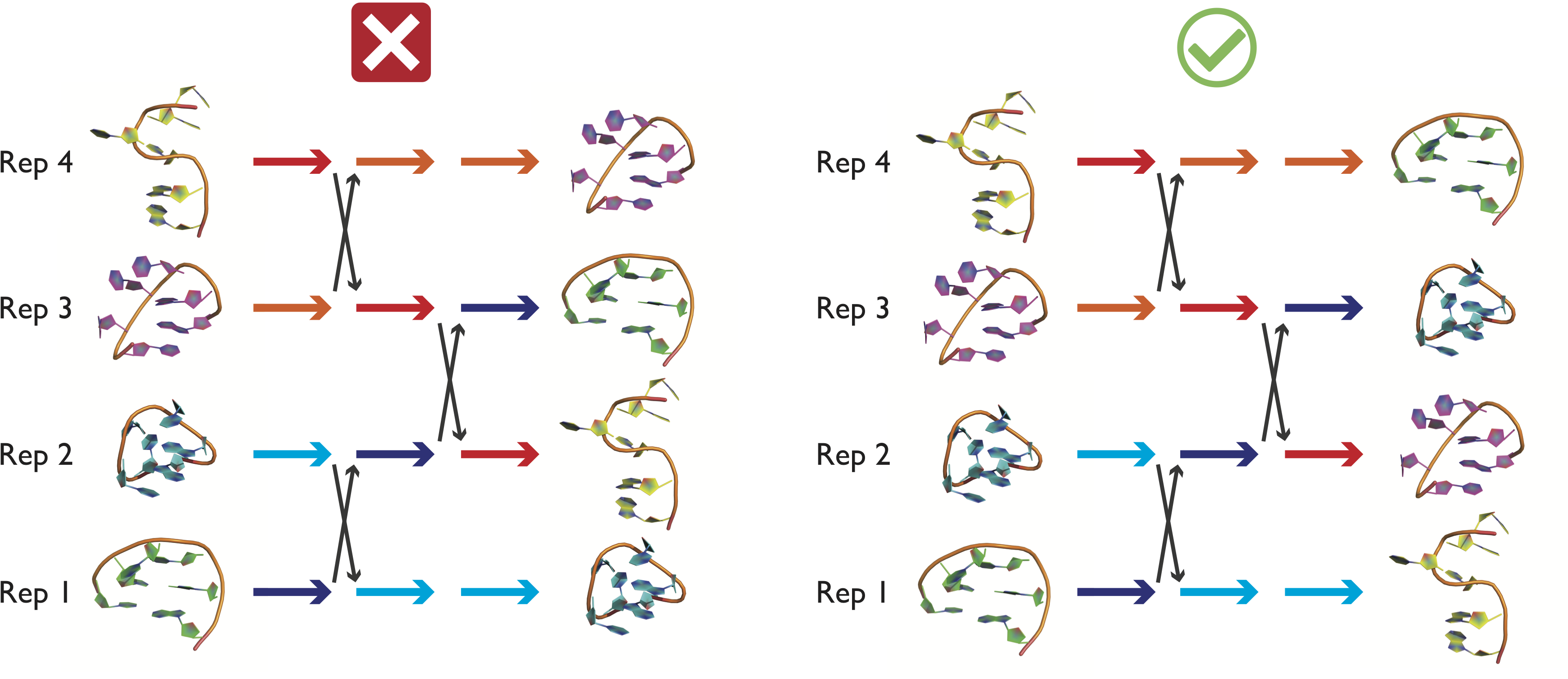}\caption{Scheme representing two possible behaviors during replica-exchange
simulations. The four different molecular structures schematically
represent four different conformations. Horizontal axis represents
time, and vertical axis the replica index. The initial and final structures
are different in each replica within both panels. However, conformational
changes illustrated in the left panel are only due to replica exchanges
(black arrows), and the continuous trajectories obtained following
the arrows with respective colors are stuck in a single conformation.
In this pathological case, the resulting populations would be affected
by a significant systematic error. This analysis applies to all replica-based
methods, including both methods based on CVs (such as replica-exchange
US) and methods based on tempering (such as T-REMD).\label{fig:Scheme-representing-two}}
\end{figure}
A popular enhanced sampling method in this community is T-REMD, probably
also thanks to its wide availability and ease of use \cite{bottaro2016rna,gil2016empirical,Kuhrova2016,bottaro2016free,yang2016predicting,gil2015,xue2015folding,park2015crystallographic,bergonzo2015stem,miner2016free,takahashi2016using,warfield2017molecular,pathak2017water,miner2017equilibrium,swadling2015structure,schuabb2017pressure}.
Other tempering methods are also notable. In particular, multi-dimensional
REMD schemes (M-REMD) where dihedral torsional potentials are in addition
scaled were successfully used for sampling unstructured oligonucleotides
\cite{bergonzo2015highly,bergonzo2015improved}, and replica exchange
with solute tempering was used to fold a tetraloop \cite{Kuhrova2016}.
Special caution should be taken into account when
analyzing replica-exchange simulations, since 
continuous trajectories should display transitions between
the relevant substates (Figure \ref{fig:Scheme-representing-two})
\cite{bottaro2016rna,Kuhrova2016,gil2015}. Many other applications
take advantage of CV-based methods. The most popular choice here is
the US method used in its implementation where multiple restraints
are used to gradually bring the system from one state to another and
the resulting trajectories are combined using the weighted-histogram
analysis method \cite{aytenfisu2017revised,di2015kissing,larsen2015thermodynamic,yildirim2015computational,wu2015multivalent,vukovic2016molecular,hayatshahi2017investigating,sun2017protonation,panteva2015force,hayatshahi2017computational,zhang2015role,thaplyal2015inverse,gaines2016ribozyme,zhang2016assessing,bingaman2017glcn6p,chen2017divalent}.
In the case of complex conformational transitions that are not sufficiently
described by the employed CVs, results of multiple-restraints US could
be highly dependent on the protocol used to initialize the simulations
\cite{di2015kissing} introducing systematic errors that are difficult
to detect. A more robust alternative is provided by replica-exchange
US simulations, used for instance in Refs. \cite{radak2015characterization,vukovic2016molecular,radak2015assessment},
at least if continuous trajectories are analyzed and transition events
are detected as discussed above for T-REMD (Figure \ref{fig:Scheme-representing-two}).
We notice that several works related to catalysis took advantage of
the string method in order to sample a reactive pathway \cite{zhang2015role,thaplyal2015inverse,zhang2016assessing,bingaman2017glcn6p}.
This approach allows multiple CVs to be combined at the price of limiting
the exploration to a reaction tube. Another popular method is metadynamics,
where an adaptive bias potential is constructed iteratively so as
to induce conformational transitions in a small number of preselected
CVs \cite{Kuhrova2016,haldar2015insights,verona2017focus,pan2017structure,bochicchio2015molecular}.
Metadynamics can be run with multiple replicas in order to accelerate
convergence \cite{mlynsky2016inline}, and can be used with a larger
number of CVs either by biasing them one-at-a-time, as in bias-exchange
metadynamics \cite{bian2015free,cunha2016ions}, or concurrently,
as in replica exchange with CV tempering \cite{cesari2016maxent,yang2016predicting,gil2015}.
Finally, metadynamics and T-REMD can be combined, as done for instance
in Refs. \cite{bottaro2016free,yang2016predicting,pathak2017water}.

\paragraph*{Employed CVs}

The success of CV-based methods depends heavily on the chosen CVs.
Many of the works discussed here used simple geometric CVs such as
distances between atoms or atom groups. Chemical reactions are typically
accomplished by biasing a combination of distances, where some describe
newly formed or expired contacts and others enforce related proton
transfers \cite{radak2015assessment,zhang2015role,zhang2016assessing,casalino2016activates,mlynsky2016inline,bingaman2017glcn6p,chen2017divalent}.
Dihedral angles can be used to enforce the exploration of multiple
rotamers \cite{gil2016empirical,cesari2016maxent,aytenfisu2017revised,gil2015,radak2015characterization,borkar2017simultaneous}.
In some cases, the barriers associated to sugar repuckering were accelerated
using pseudodihedrals \cite{gil2016empirical,cesari2016maxent,yang2016predicting,gil2015,radak2015characterization}.
Some other works used root-mean-square deviation (RMSD) from the native
structure \cite{Kuhrova2016,haldar2015insights,vukovic2016molecular,palermo2017crispr}.
RMSD is known to be a poor descriptor, in particular for RNA systems.
Refs. \cite{bottaro2016free,yang2016predicting} report cases where
an RNA-dedicated metric ($\epsilon$RMSD) was utilized to distinguish
native from non-native structures and biased.
Whereas measures such as the $\epsilon$RMSD distinguish structures
using the entire map of observed and non-observed contacts,
variables such as the number of native contacts
\cite{Kuhrova2016,haldar2015insights,borkar2016structure,borkar2017simultaneous,pathak2017water}
are unaffected by the presence
or absence of competing non-native contacts. As a general consideration,
it should be taken into account that, for intrinsically high-dimensional
free-energy landscapes, identifying a small number of CVs capable
to describe all the relevant substates might be difficult or even
virtually impossible.

\section*{Discussion and Perspectives}

In this Review we surveyed the enhanced sampling methods recently
applied to study RNA structural dynamics. In the following, we summarize
our recommendations and perspectives.

\paragraph*{Different Methods for the Same Problem}

Different groups used different methods to tackle very similar problems.
An example can be seen by comparing three works where the affinities
of divalent ions for specific sites in RNA motifs were computed, ranging
from classical US \cite{panteva2015force} through a novel grand-canonical
Monte Carlo/MD approach \cite{lemkul2016characterization} to a bias-exchange-like
metadynamics procedure \cite{cunha2016ions}. It would be interesting
to compare these three methods on identical setups in order to assess
their computational efficiency. Similarly, related catalytic reactions
were tackled by different authors using US combined with string method
\cite{zhang2015role,thaplyal2015inverse,zhang2016assessing,bingaman2017glcn6p} or thermodynamic
integration (TI) \cite{casalino2016activates}. Albeit employed on
different systems, both approaches used comparable QM/MM setups with
similar sized QM regions described by the same DFT functional. String
method calculations might allow a more precise definition of TSs,
although the employed iterative procedure is expensive. The TI approach
exploits a monodimensional pathway and required a verification that
the related proton transfer was in fact induced in a reversible manner.
Whereas the rearrangements associated to phosphodiester cleavage reactions
are significantly simpler than those associated to RNA folding, we
suggest that replica-exchange procedures where coordinates are swapped
between consecutive windows might be beneficial in performing cleavage
simulations, allowing reactive events to be observed in continuous
trajectories.

\paragraph*{Recommendations for Enhanced Sampling Simulations}

At variance with plain MD simulations, which are often analyzed in
a qualitative fashion, enhanced sampling simulations are usually employed
to report thermodynamic averages to be directly compared with experiments.
For a quantitative interpretation, it is however necessary that statistical
errors are reported together with the results. Since the estimate
of the error itself is sometimes non trivial, we suggest that authors
should explain clearly how the errors were computed. We would like
to reiterate that enhanced sampling methods can give statistically
reliable results only if multiple transitions are observed in continuous
trajectories. In addition, given the difficulty in reproducing this
kind of calculations, authors should share input parameters
and, when feasible, generated trajectories.
For instance, the protocols introduced in \cite{bottaro2016free} and \cite{gil2015}
were used after a very short time by an independent group \cite{yang2016predicting}.
Finally, CV-based
methods usually require a significant number of trial and errors in
order to identify proper CVs. Sharing the non-working setups, perhaps
in the form of supporting information, could speed up the progress
in the field avoiding other groups to repeat similar mistakes.

\paragraph*{Recommendations for RNA Simulations}

In the last two years, mostly thanks to the publication of extensive
benchmarks using enhanced sampling techniques, it was suggested that
current empirical force fields are not yet accurate enough for blind prediction
of some RNA native structures and for reliably reproducing the conformational
ensembles of small unstructured RNAs. Nevertheless, even without predictive accuracy,
MD simulations are able to provide significant insights on experiments,
mostly thanks to their spatial and temporal resolution. However, we
feel that the community should work in the direction of improving
force fields, taking advantage of enhanced sampling techniques in
order to validate them. In this respect, we believe it is crucial
that researchers continue sharing benchmarks and negative results.
In addition, predictive simulations
should be validated with care against independent experimental data.
In this respect, solution-phase experiments such as NMR are particularly useful since they provide
ensemble averages that can be directly compared with MD simulations and that account for RNA dynamics.
A promising growing field is based on the idea of simultaneously applying
enhanced sampling simulations and restraints obtained from experiments
\cite{cesari2016maxent,borkar2016structure,borkar2017simultaneous}.

\paragraph*{Perspectives}

The importance of RNA structural dynamics in molecular biology is
steadily growing. Structures of new RNA catalytic systems are being
discovered at a constant pace, and hypotheses on their mechanism of
action benefit from explicit modeling of the corresponding reaction
pathways. In addition, local dynamics of flexible RNA motifs, especially
in relation to their capability to bind ions, small ligands, proteins,
and other RNA molecules, is receiving an increasing attention.
We thus predict
the role of enhanced sampling techniques in the RNA community to increase in the
next years.

\section*{Acknowledgments}

Alejandro Gil-Ley, Sandro Bottaro, Carlo Camilloni, Angel E. Garcia, Alex MacKerell,
and Alessandra Magistrato are acknowledged for reading the manuscript
and providing useful suggestions. This work has been supported European
Research Council under the European Union\textquoteright s Seventh
Framework Programme (FP/2007-2013)/ERC Grant Agreement n. 306662,
S-RNA-S.

\bibliographystyle{elsarticle-num}
\bibliography{bibliography}

\end{document}

%% file: tableshort.tex
\begin{tabular}{cccccc}
\toprule 
Enhanced Sampling Method   &  CVs  &  RNA Systems  & QM/MM &  Total Timescale ($\mu$s)  &  Ref.\tabularnewline
\midrule
\midrule 
\multicolumn{6}{l}{\emph{Refinements and Validations of Force Fields}}\tabularnewline
\midrule 
M-REMD  &  -  &  UUCG-TL (10,14-mers), CCCC, GACC  & MM &  $\sim$7663  &  \cite{bergonzo2015highly}\tabularnewline
\midrule 
M-REMD  &  -  &  CCCC, GACC  & MM &  $\sim$1382.4  &  \cite{bergonzo2015improved}\tabularnewline
\midrule 
H-REMD  &  -  &  single strands (6-mers), duplexes (12-mers)  & MM &  1.68  &  \cite{sakuraba2015predicting}\tabularnewline
\midrule 
T-REMD  &  -  &  AAAA, CAAU, CCCC, GACC, UUUU  & MM &  264  &  \cite{bottaro2016rna}\tabularnewline
\midrule 
METAD, T-REMD, RECT  &  $\alpha$,$\beta$,$\gamma$,$\epsilon$,$\zeta$,$Z_{x}$,$\chi$, distance (COM)  &  CC, AA, CA, AC, GACC, CCCC, AAAA  & MM &  $\sim$206  &  \cite{gil2016empirical}\tabularnewline
\midrule 
METAD, T-REMD, REST2  &  H-bonds, RMSD   &  GAGA-TL (8,10-mers)  & MM &  966  &  \cite{Kuhrova2016}\tabularnewline
\midrule 
T-REMD+METAD  &  $\epsilon$RMSD  &  GAGA-TL, UUCG-TL (6,8-mers)  & MM &  96  &  \cite{bottaro2016free}\tabularnewline
\midrule 
RECT  &  $\alpha$,$\beta$,$\gamma$,$\epsilon$,$\zeta$,$Z_{x}$,$Z_{y}$,$\chi$, distance (COM)  &  A, C, AA, AC, CA, CC  & MM &  $\sim$35  &  \cite{cesari2016maxent}\tabularnewline
\midrule 
RECT, T-REMD+METAD  &  $\alpha$,$\beta$,$\gamma$,$\epsilon$,$\zeta$,$Z_{x}$,$Z_{y}$,$\chi$, distance (COM), $\epsilon$RMSD  &  GACC, CCCC, AAAA, CAAA, UUCG-TL (8-mer)   & MM &  $\sim$504  &  \cite{yang2016predicting}\tabularnewline
\midrule 
US  &  $\alpha$,$\beta$,$\gamma$,$\epsilon$,$\zeta$,$\chi$  &  16 dinucleotides  & MM &  16.128  &  \cite{aytenfisu2017revised}\tabularnewline
\midrule 
\multicolumn{6}{l}{\emph{Conformational Landscapes}}\tabularnewline
\midrule 
RECT, H-REMD, T-REMD  &  $\alpha$,$\beta$,$\gamma$,$\delta$,$\epsilon$,$\zeta$,$Z_{x}$,$Z_{y}$,$\chi$, distance (COM)  &  GACC  & MM &  $\sim$14.4  &  \cite{gil2015}\tabularnewline
\midrule 
METAD  &  H-bonds, RGyr, RMSD, $\chi$  &  GAGA-TL, UUCG-TL (10-mers)  & MM &  4.44  &  \cite{haldar2015insights}\tabularnewline
\midrule 
T-REMD  &  -  &  SAM-II riboswitch  & MM &  6  &  \cite{xue2015folding}\tabularnewline
\midrule 
T-REMD, SMD  &  distance (COM)  &  (cgauUCUaugc) duplex (22-mer)  & MM &  $\sim$5  &  \cite{park2015crystallographic}\tabularnewline
\midrule 
T-REMD  &  -  &  SVL loop (17-mer)  & MM &  57.6  &  \cite{bergonzo2015stem}\tabularnewline
\midrule 
A-REMD  &  $Z_{x}$, $Z_{y}$, $\chi$  &  U nucleoside  & QM/MM &  $\sim$0.1  &  \cite{radak2015characterization}\tabularnewline
\midrule 
BE-METAD  &  H-bonds, RGyr, energy  &  gene32 mRNA pseudoknot (32-mer)  & MM &  3  &  \cite{bian2015free}\tabularnewline
\midrule 
US  &  distance (COM)  &  \emph{add }riboswitch  & MM &  1.177  &  \cite{di2015kissing}\tabularnewline
\midrule 
US  &  $\chi$  &  U and 2-thio-U nucleosides  & MM &  0.144  &  \cite{larsen2015thermodynamic}\tabularnewline
\midrule 
US  &  $\chi$, pseudodihedral (COM)  &  A, G, U, and C nucleosides, duplex with CUG (18-mer)  & MM &  6.016  &  \cite{yildirim2015computational}\tabularnewline
\midrule
US+pseudo-spring method  &  distance (COM)  &  duplex (32-mer)  & MM &  $\sim$0.3  &  \cite{wu2015multivalent}\tabularnewline
\midrule 
T-REMD  &  -  &  GCAA -TL (8-mer)  & MM &  448   &  \cite{miner2016free}\tabularnewline
\midrule
T-REMD  &  -  &  pT181 RNA hairpins (48-mers)  & MM &  17.16  &  \cite{takahashi2016using}\tabularnewline
\midrule
RAM  &  H-bonds, distance (COM)  &  TAR (29-mer) in RNA:peptide complex  & MM &  0.8  &  \cite{borkar2016structure}\tabularnewline
\midrule 
REMD+US  &  distance, RMSD  &  U-singlestrand (5-mer) in RNA:protein complex  & MM &  3.6  &  \cite{vukovic2016molecular}\tabularnewline
\midrule 
US  &  distance (COM)  &  GTPase center of rRNA  & MM &  4.494  &  \cite{hayatshahi2017investigating}\tabularnewline
\midrule
RAM  &  $\alpha$,$\beta$,$\delta$,$\epsilon$,$\zeta$, H-bonds   &  UUCG-TL (14-mer)  & MM &  1.04  &  \cite{borkar2017simultaneous}\tabularnewline
\midrule
T-REMD  &  -  &  theophylline-binding aptamer (33-mer)  & MM &  1.6  &  \cite{warfield2017molecular}\tabularnewline
\midrule 
GaMD, TMD  &  RMSD  &  CRISPR-Cas9 RNA complex  & MM &  $\sim$12  &  \cite{palermo2017crispr}\tabularnewline
\midrule
METAD  &  distance (COM), stacking  &  PNAs (6-mers), PNA:RNA duplex (12-mer)  & MM &  $\sim$1.2  &  \cite{verona2017focus}\tabularnewline
\midrule
T-REMD+METAD  &  H-bonds, RGyr  &  polio viral RNA hairpin (22-mer)  & MM &  16  &  \cite{pathak2017water}\tabularnewline
\midrule
US  &  pseudodihedral (COM)  &  hairpin from group-II intron (35-mer)  & MM &  1.008  &  \cite{sun2017protonation}\tabularnewline
\midrule 
METAD  &  $\chi$, pseudodihedral (COM)  &  duplexes with A-A mismatches (18-mers)  & MM &  $\sim$0.6  &  \cite{pan2017structure}\tabularnewline
\midrule
T-REMD  &  -  &  gcGCAAgc-TL (8-mer)  & MM &  356  &  \cite{miner2017equilibrium}\tabularnewline
\midrule
\multicolumn{6}{l}{\emph{Ion Interactions and Ion/ligand Induced Conformational Changes}}\tabularnewline
\midrule 
METAD  &  distance (COM), hydrophobic contacts, H-bonds  &  duplex with A-A mismatches (20-mer)   & MM &  1.4  &  \cite{bochicchio2015molecular}\tabularnewline
\midrule
US+TI  &  distance  &  mononucleotides, hammerhead ribozyme  & MM &  $\sim$3.5  &  \cite{panteva2015force}\tabularnewline
\midrule 
SMD  &  distance  &  preQ$_{1}$-III riboswitch  & MM &  0.1  &  \cite{liberman2015structural}\tabularnewline
\midrule 
BE-METAD  &  distance, coordination  &  nucleosides, GpG dinucleotide, GC duplex (8-mer)  & MM &  82  &  \cite{cunha2016ions}\tabularnewline
\midrule 
GCMC-MD  &  distance, coordination  &  BWYV pseudoknot, VS ribozyme, 23S rRNA, Mg$^{2+}$ riboswitch  & MM &  4  &  \cite{lemkul2016characterization}\tabularnewline
\midrule 
US  &  distance (COM)  &  GTPase center of rRNA  & MM &  20.488  &  \cite{hayatshahi2017computational}\tabularnewline
\midrule 
TI  &  distance  &  guanine riboswitch  & MM &  0.576  &  \cite{hu2017ligand}\tabularnewline
\midrule 
T-REMD, METAD  &  distance (COM)  &  siRNA duplex (42-mer)  & MM &  28.8  &  \cite{grasso2017free}\tabularnewline
\midrule
\multicolumn{6}{l}{\emph{Reactivity and Catalysis}}\tabularnewline
\midrule 
A-REMD  &  distance  &  HDV ribozyme  & QM/MM &  $\sim$0.235  &  \cite{radak2015assessment}\tabularnewline
\midrule 
US+string method  &  distance  &  \emph{glmS} riboswitch  & QM/MM &  0.00225, $\sim$0.00015  &  \cite{zhang2015role}\tabularnewline
\midrule 
T-REMD  &  -  &  hammerhead ribozyme  & MM &  5  &  \cite{swadling2015structure}\tabularnewline
\midrule 
US+string method  &  distance, angle  &  HDV ribozyme  & MM, QM/MM  &  0.0465, $\sim$0.00019  &  \cite{thaplyal2015inverse}\tabularnewline
\midrule 
US, TI  &  distance  &  twister ribozyme  & MM &  0.525  &  \cite{gaines2016ribozyme}\tabularnewline
\midrule 
US+string method  &  distance  &  \emph{glmS} riboswitch  & QM/MM &  $\sim$0.000098  &  \cite{zhang2016assessing}\tabularnewline
\midrule 
TI  &  distance  &  group-II introns  & QM/MM &  $\sim$0.0001  &  \cite{casalino2016activates}\tabularnewline
\midrule 
METAD  &  distance  &  GAAA-TL, UUCG-TL (10-mers), GC duplex (16-mer)  & QM/MM &  0.72  &  \cite{mlynsky2016inline}\tabularnewline
\midrule 
US+string method  &  distance  &  \emph{glmS} riboswitch  & QM/MM &  $\sim$0.0072  &  \cite{bingaman2017glcn6p}\tabularnewline
\midrule 
T-REMD  &  -  &  hairpin ribozyme  & MM &  25.6  &  \cite{schuabb2017pressure}\tabularnewline
\midrule 
US, TI  &  distance  &  hammerhead ribozyme  & MM, QM/MM  &  $\sim$0.33, $\sim$0.0004  &  \cite{chen2017divalent}\tabularnewline
\bottomrule
\end{tabular}


%% file: manuscript_arxiv.bbl
\begin{thebibliography}{10}
\expandafter\ifx\csname url\endcsname\relax
  \def\url#1{\texttt{#1}}\fi
\expandafter\ifx\csname urlprefix\endcsname\relax\def\urlprefix{URL}\fi
\expandafter\ifx\csname href\endcsname\relax
  \def\href#1#2{#2} \def\path#1{#1}\fi

\bibitem{lilley2017rna}
D.~M. Lilley, How {RNA} acts as a nuclease: {S}ome mechanistic comparisons in
  the nucleolytic ribozymes, Biochem. Soc. Trans. 45~(3) (2017) 683--691.

\bibitem{morris2014riseregRNA}
K.~V. Morris, J.~S. Mattick, The rise of regulatory {RNA}, Nat. Rev. Genet.
  15~(6) (2014) 423.

\bibitem{westhof2015twenty}
E.~Westhof, Twenty years of {RNA} crystallography, RNA 21~(4) (2015) 486--487.

\bibitem{rinnenthal2011mapping}
J.~Rinnenthal, J.~Buck, J.~Ferner, A.~Wacker, B.~F{{\"u}}rtig, H.~Schwalbe,
  Mapping the landscape of {RNA} dynamics with {NMR} spectroscopy, Acc. Chem.
  Res. 44~(12) (2011) 1292--1301.

\bibitem{lilley2012structure}
D.~M. Lilley, The structure and folding of kink turns in {RNA}, Wiley
  Interdiscip. Rev. RNA 3~(6) (2012) 797--805.

\bibitem{dethoff2012visualizing}
E.~A. Dethoff, K.~Petzold, J.~Chugh, A.~Casiano-Negroni, H.~M. Al-Hashimi,
  Visualizing transient low-populated structures of {RNA}, Nature 491~(7426)
  (2012) 724--728.

\bibitem{serganov2013decade}
A.~Serganov, E.~Nudler, A decade of riboswitches, Cell 152~(1) (2013) 17--24.

\bibitem{emilsson2003ribozyme}
G.~M. Emilsson, S.~Nakamura, A.~Roth, R.~R. Breaker, Ribozyme speed limits, RNA
  9~(8) (2003) 907--918.

\bibitem{pan2016demonstrating}
A.~C. Pan, T.~M. Weinreich, S.~Piana, D.~E. Shaw, Demonstrating an
  order-of-magnitude sampling enhancement in molecular dynamics simulations of
  complex protein systems, J. Chem. Theory Comput. 13~(3) (2016) 1360--1367.

\bibitem{bernardi2015enhanced}
R.~C. Bernardi, M.~C. Melo, K.~Schulten, Enhanced sampling techniques in
  molecular dynamics simulations of biological systems, Biochim. Biophys. Acta,
  Gen. Subj. 1850~(5) (2015) 872--877.

\bibitem{valsson2016enhancing}
O.~Valsson, P.~Tiwary, M.~Parrinello, Enhancing important fluctuations: {R}are
  events and metadynamics from a conceptual viewpoint, Annu. Rev. Phys. Chem.
  67 (2016) 159--184.

\bibitem{vsponer2013understand}
J.~{\v{S}}poner, J.~E. {\v{S}}poner, A.~Ml{\'a}dek, P.~Ban{\'a}{\v{s}},
  P.~Jure{\v{c}}ka, M.~Otyepka, How to understand quantum chemical computations
  on {DNA} and {RNA} systems? {A} practical guide for non-specialists, Methods
  64~(1) (2013) 3--11.

\bibitem{huang2015nucleic}
M.~Huang, T.~J. Giese, D.~M. York, Nucleic acid reactivity: {C}hallenges for
  next-generation semiempirical quantum models, J. Comput. Chem. 36~(18) (2015)
  1370--1389.

\bibitem{vsponer2017understand}
J.~{\v{S}}poner, M.~Krepl, P.~Ban{\'a}{\v{s}}, P.~K{\"u}hrov{\'a},
  M.~Zgarbov{\'a}, P.~Jure{\v{c}}ka, M.~Havrila, M.~Otyepka, How to understand
  atomistic molecular dynamics simulations of {RNA} and protein--{RNA}
  complexes?, Wiley Interdiscip. Rev. RNA 8~(3) (2017) e1405.

\bibitem{vangaveti2017advances}
S.~Vangaveti, S.~V. Ranganathan, A.~A. Chen, Advances in {RNA} molecular
  dynamics: {A} simulator's guide to {RNA} force fields, Wiley Interdiscip.
  Rev. RNA 8~(2) (2017) e1396.

\bibitem{smith2017physics}
L.~G. Smith, J.~Zhao, D.~H. Mathews, D.~H. Turner, Physics-based all-atom
  modeling of {RNA} energetics and structure, Wiley Interdiscip. Rev. RNA 8~(5)
  (2017) e1422.

\bibitem{bergonzo2015highly}
C.~Bergonzo, N.~M. Henriksen, D.~R. Roe, T.~E. Cheatham~III, Highly sampled
  tetranucleotide and tetraloop motifs enable evaluation of common {RNA} force
  fields, RNA 21~(9) (2015) 1578--1590, { \\$^{\bullet{}\bullet{}}$ Extensive
  investigation on several tetraloops and tetranucleotides using a
  multidimensional replica-exchange scheme. For tetranucleotides, this work
  provides the first virtually converged simulation, showing that the
  population of intercalated structures is larger than expected from {NMR} data
 }.

\bibitem{bergonzo2015improved}
C.~Bergonzo, T.~E. Cheatham~III, Improved force field parameters lead to a
  better description of {RNA} structure, J. Chem. Theory Comput. 11~(9) (2015)
  3969--3972.

\bibitem{sakuraba2015predicting}
S.~Sakuraba, K.~Asai, T.~Kameda, Predicting {RNA} duplex dimerization
  free-energy changes upon mutations using molecular dynamics simulations, J.
  Phys. Chem. Lett. 6~(21) (2015) 4348--4351.

\bibitem{bottaro2016rna}
S.~Bottaro, A.~Gil-Ley, G.~Bussi, {{RNA}} folding pathways in stop motion,
  Nucleic Acids Res. 44~(12) (2016) 5883--5891.

\bibitem{gil2016empirical}
A.~Gil-Ley, S.~Bottaro, G.~Bussi, Empirical corrections to the {AMBER} {RNA}
  force field with target metadynamics, J. Chem. Theory Comput. 12~(6) (2016)
  2790--2798.

\bibitem{Kuhrova2016}
P.~K{\"u}hrov{\'a}, R.~B. Best, S.~Bottaro, G.~Bussi, J.~{\v{S}}poner,
  M.~Otyepka, P.~Ban{\'a}{\v{s}}, Computer folding of {RNA} tetraloops:
  {I}dentification of key force field deficiencies, J. Chem. Theory Comput.
  12~(9) (2016) 4534--4548, { \\$^{\bullet{}\bullet{}}$ Extensive investigation
  on a tetraloop using several enhanced sampling methods including temperature
  replica exchange, solute tempering, and metadynamics. Different methods
  result in qualitatively consistent results. The most important current force
  fields are compared, and none of them can reproduce a significant population
  of the known native structure}.

\bibitem{bottaro2016free}
S.~Bottaro, P.~Ban{\'a}{\v s}, J.~{{\v S}}poner, G.~Bussi, Free energy
  landscape of {{GAGA}} and {{UUCG}} {{RNA}} tetraloops, J. Phys. Chem. Lett.
  7~(20) (2016) 4032--4038, { \\$^{\bullet{}\bullet{}}$ Converged free-energy
  landscapes for two {RNA} tetraloops including both native and non-native
  structures are here reported for the first time. The result is obtained
  exploiting a {RNA}-specific structural metrics as biased collective variable.
  This procedure allows to quantify the (in)stability of the native structure.
  In addition, a direct comparison with available NMR data is performed}.

\bibitem{cesari2016maxent}
A.~Cesari, A.~Gil-Ley, G.~Bussi, Combining simulations and solution experiments
  as a paradigm for {RNA} force field refinement, J. Chem. Theory Comput.
  12~(12) (2016) 6192--6200.

\bibitem{yang2016predicting}
C.~Yang, M.~Lim, E.~Kim, Y.~Pak, Predicting {RNA} structures via a simple van
  der {W}aals correction to an all-atom force field, J. Chem. Theory Comput.
  13~(2) (2017) 395--399.

\bibitem{aytenfisu2017revised}
A.~H. Aytenfisu, A.~Spasic, A.~Grossfield, H.~A. Stern, D.~H. Mathews, Revised
  {RNA} dihedral parameters for the {AMBER} force field improve {RNA} molecular
  dynamics, J. Chem. Theory Comput. 13~(2) (2017) 900--915.

\bibitem{gil2015}
A.~Gil-Ley, G.~Bussi, Enhanced conformational sampling using replica exchange
  with collective-variable tempering, J. Chem. Theory Comput. 11~(3) (2015)
  1077--1085, { \\$^{\bullet{}\bullet{}}$ A novel enhanced sampling method is
  introduced that combines metadynamics and replica exchange. The method
  accelerates simultaneously many collective variables and is particularly
  suitable to accelerate transitions between different {RNA} rotamers}.

\bibitem{haldar2015insights}
S.~Haldar, P.~K{\"u}hrov{\'a}, P.~Ban{\'a}{\v{s}}, V.~Spiwok, J.~{\v{S}}poner,
  P.~Hobza, M.~Otyepka, Insights into stability and folding of {GNRA} and
  {UNCG} tetraloops revealed by microsecond molecular dynamics and
  well-tempered metadynamics., J. Chem. Theory Comput. 11~(8) (2015) 3866.

\bibitem{xue2015folding}
X.~Xue, W.~Yongjun, L.~Zhihong, Folding of {SAM-II} riboswitch explored by
  replica-exchange molecular dynamics simulation, J. Theor. Biol. 365 (2015)
  265--269.

\bibitem{park2015crystallographic}
H.~Park, {\`A}.~L. Gonz{\'a}lez, I.~Yildirim, T.~Tran, J.~R. Lohman, P.~Fang,
  M.~Guo, M.~D. Disney, Crystallographic and computational analyses of {AUUCU}
  repeating {RNA} that causes {S}pinocerebellar {A}taxia type 10 ({SCA10}),
  Biochemistry 54~(24) (2015) 3851.

\bibitem{bergonzo2015stem}
C.~Bergonzo, K.~B. Hall, T.~E. Cheatham~III, Stem-{L}oop {V} of {V}arkud
  {S}atellite {RNA} exhibits characteristics of the {Mg}$^{2+}$ bound structure
  in the presence of monovalent ions, J. Phys. Chem. B 119~(38) (2015)
  12355--12364.

\bibitem{radak2015characterization}
B.~K. Radak, M.~Romanus, T.-S. Lee, H.~Chen, M.~Huang, A.~Treikalis,
  V.~Balasubramanian, S.~Jha, D.~M. York, Characterization of the
  three-dimensional free energy manifold for the uracil ribonucleoside from
  asynchronous replica exchange simulations, J. Chem. Theory Comput. 11~(2)
  (2015) 373--377.

\bibitem{bian2015free}
Y.~Bian, J.~Zhang, J.~Wang, J.~Wang, W.~Wang, Free energy landscape and
  multiple folding pathways of an {H}-type {RNA} pseudoknot, PloS One 10~(6)
  (2015) e0129089.

\bibitem{di2015kissing}
F.~Di~Palma, S.~Bottaro, G.~Bussi, Kissing loop interaction in adenine
  riboswitch: {I}nsights from umbrella sampling simulations, BMC Bioinformatics
  16~(Suppl 9) (2015) S6, { \\$^{\bullet{}\bullet{}}$ Umbrella sampling is used
  to investigate the ligand-dependent stabilization of a kissing loop in the
  adenine riboswitch, obtaining a qualitative agreement with experiment.
  Importantly, the dependence of the results on the initialization protocol is
  discussed in detail}.

\bibitem{larsen2015thermodynamic}
A.~T. Larsen, A.~C. Fahrenbach, J.~Sheng, J.~Pian, J.~W. Szostak, Thermodynamic
  insights into 2-thiouridine-enhanced {RNA} hybridization, Nucleic Acids Res.
  43~(16) (2015) 7675--7687.

\bibitem{yildirim2015computational}
I.~Yildirim, D.~Chakraborty, M.~D. Disney, D.~J. Wales, G.~C. Schatz,
  Computational investigation of {RNA} {CUG} repeats responsible for myotonic
  dystrophy 1, J. Chem. Theory Comput. 11~(10) (2015) 4943--4958.

\bibitem{wu2015multivalent}
Y.-Y. Wu, Z.-L. Zhang, J.-S. Zhang, X.-L. Zhu, Z.-J. Tan, Multivalent
  ion-mediated nucleic acid helix-helix interactions: {RNA} versus {DNA},
  Nucleic Acids Res. 43~(12) (2015) 6156--6165.

\bibitem{miner2016free}
J.~C. Miner, A.~A. Chen, A.~E. Garc{\'\i}a, Free-energy landscape of a
  hyperstable {RNA} tetraloop, Proc. Natl. Acad. Sci. U.S.A. 113~(24) (2016)
  6665--6670.

\bibitem{takahashi2016using}
M.~K. Takahashi, K.~E. Watters, P.~M. Gasper, T.~R. Abbott, P.~D. Carlson,
  A.~A. Chen, J.~B. Lucks, Using in-cell {SHAPE}-seq and simulations to probe
  structure--function design principles of {RNA} transcriptional regulators,
  RNA 22~(6) (2016) 920--933.

\bibitem{borkar2016structure}
A.~N. Borkar, M.~F. Bardaro, C.~Camilloni, F.~A. Aprile, G.~Varani,
  M.~Vendruscolo, Structure of a low-population binding intermediate in
  protein--{RNA} recognition, Proc. Natl. Acad. Sci. U.S.A. 113~(26) (2016)
  7171--7176, { \\$^{\bullet{}\bullet{}}$ The structural dynamics of an
  {RNA}:peptide complex formed by {TAR} {RNA} and a cyclic peptide is
  characterized. By combining enhanced sampling simulations with {NMR} data the
  authors are able to identify low population structures of the complex that
  are then validated experimentally. This work shows how to effectively combine
  experimental data and enhanced sampling simulations}.

\bibitem{vukovic2016molecular}
L.~Vukovic, C.~Chipot, D.~L. Makino, E.~Conti, K.~Schulten, Molecular mechanism
  of processive 3' to 5' {RNA} translocation in the active subunit of the {RNA}
  exosome complex, J. Am. Chem. Soc. 138~(12) (2016) 4069--4078.

\bibitem{hayatshahi2017investigating}
H.~S. Hayatshahi, C.~Bergonzo, T.~E. Cheatham~III, Investigating the ion
  dependence of the first unfolding step of {GTP}ase-associating center
  ribosomal {RNA}, J. Biomol. Struct. Dyn. (2017) 1--11.

\bibitem{borkar2017simultaneous}
A.~N. Borkar, P.~Vallurupalli, C.~Camilloni, L.~E. Kay, M.~Vendruscolo,
  Simultaneous {NMR} characterisation of multiple minima in the free energy
  landscape of an {RNA} {UUCG} tetraloop, Phys. Chem. Chem. Phys. 19~(4) (2017)
  2797--2804.

\bibitem{warfield2017molecular}
B.~M. Warfield, P.~C. Anderson, Molecular simulations and {M}arkov state
  modeling reveal the structural diversity and dynamics of a
  theophylline-binding {RNA} aptamer in its unbound state, PloS One 12~(4)
  (2017) e0176229.

\bibitem{palermo2017crispr}
G.~Palermo, Y.~Miao, R.~C. Walker, M.~Jinek, J.~A. McCammon, {CRISPR}-{C}as9
  conformational activation as elucidated from enhanced molecular simulations,
  Proc. Natl. Acad. Sci. U.S.A. 114~(28) (2017) 7260--7265.

\bibitem{verona2017focus}
M.~D. Verona, V.~Verdolino, F.~Palazzesi, R.~Corradini, Focus on {PNA}
  flexibility and {RNA} binding using molecular dynamics and metadynamics, Sci.
  Rep. 7 (2017) 42799.

\bibitem{pathak2017water}
A.~K. Pathak, T.~Bandyopadhyay, Water isotope effect on the thermostability of
  a polio viral {RNA} hairpin: {A} metadynamics study, J. Chem. Phys. 146~(16)
  (2017) 165104.

\bibitem{sun2017protonation}
Z.~Sun, X.~Wang, J.~Z. Zhang, Protonation-dependent base flipping in the
  catalytic triad of a small {RNA}, Chem. Phys. Lett. 684 (2017) 239--244.

\bibitem{pan2017structure}
F.~Pan, V.~H. Man, C.~Roland, C.~Sagui, Structure and dynamics of {DNA} and
  {RNA} double helices of {CAG} and {GAC} trinucleotide repeats, Biophys. J.
  113~(1) (2017) 19--36.

\bibitem{miner2017equilibrium}
J.~C. Miner, A.~E. Garc{\'\i}a, Equilibrium denaturation and preferential
  interactions of an {RNA} tetraloop with urea, J. Phys. Chem. B 121~(15)
  (2017) 3734--3746.

\bibitem{bochicchio2015molecular}
A.~Bochicchio, G.~Rossetti, O.~Tabarrini, S.~Krau$\beta$, P.~Carloni, Molecular
  view of ligands specificity for {CAG} repeats in anti-{H}untington therapy,
  J. Chem. Theory Comput. 11~(10) (2015) 4911--4922.

\bibitem{panteva2015force}
M.~T. Panteva, G.~M. Giambasu, D.~M. York, Force field for {Mg}$^{2+}$,
  {Mn}$^{2+}$, {Zn}$^{2+}$, and {Cd}$^{2+}$ ions that have balanced
  interactions with nucleic acids, J. Phys. Chem. B 119~(50) (2015)
  15460--15470.

\bibitem{liberman2015structural}
J.~A. Liberman, K.~C. Suddala, A.~Aytenfisu, D.~Chan, I.~A. Belashov, M.~Salim,
  D.~H. Mathews, R.~C. Spitale, N.~G. Walter, J.~E. Wedekind, Structural
  analysis of a class {III} pre{Q}1 riboswitch reveals an aptamer distant from
  a ribosome-binding site regulated by fast dynamics, Proc. Natl. Acad. Sci.
  U.S.A. 112~(27) (2015) E3485--E3494.

\bibitem{cunha2016ions}
R.~A. Cunha, G.~Bussi, Unraveling {Mg}$^{2+}$--{RNA} binding with atomistic
  molecular dynamics, RNA 23~(5) (2017) 628--638.

\bibitem{lemkul2016characterization}
J.~A. Lemkul, S.~K. Lakkaraju, A.~D. MacKerell~Jr, Characterization of
  {Mg}$^{2+}$ distributions around {RNA} in solution, ACS omega 1~(4) (2016)
  680--688.

\bibitem{hayatshahi2017computational}
H.~S. Hayatshahi, D.~R. Roe, R.~Galindo-Murillo, K.~B. Hall, T.~E.
  Cheatham~III, Computational assessment of potassium and magnesium ion binding
  to a buried pocket in {GTP}ase-associating center {RNA}, J. Phys. Chem. B
  121~(3) (2017) 451--462.

\bibitem{hu2017ligand}
G.~Hu, A.~Ma, J.~Wang, Ligand selectivity mechanism and conformational changes
  in guanine riboswitch by molecular dynamics simulations and free energy
  calculations, J. Chem. Inf. Model. 57~(4) (2017) 918--928.

\bibitem{grasso2017free}
G.~Grasso, M.~A. Deriu, V.~Patrulea, G.~Borchard, M.~M{\"o}ller, A.~Danani,
  Free energy landscape of si{RNA}-polycation complexation: {E}lucidating the
  effect of molecular geometry, polymer flexibility, and charge neutralization,
  PloS One 12~(10) (2017) e0186816.

\bibitem{radak2015assessment}
B.~K. Radak, T.-S. Lee, M.~E. Harris, D.~M. York, Assessment of metal-assisted
  nucleophile activation in the hepatitis delta virus ribozyme from molecular
  simulation and {3D}-{RISM}, RNA 21~(9) (2015) 1566--1577.

\bibitem{zhang2015role}
S.~Zhang, A.~Ganguly, P.~Goyal, J.~L. Bingaman, P.~C. Bevilacqua,
  S.~Hammes-Schiffer, Role of the active site guanine in the {\it {glms}}
  ribozyme self-cleavage mechanism: {Q}uantum mechanical/molecular mechanical
  free energy simulations, J. Am. Chem. Soc. 137~(2) (2015) 784--798, {
  \\$^{\bullet{}\bullet{}}$ {QM/MM} free-energy calculations are here used to
  compare various mechanisms of phospodiester cleavage. The main focus is on
  the role of guanine nucleotide, which is present in the active sites of
  several small self-cleaving ribozymes. The effective usage of string method
  is also described in detail}.

\bibitem{swadling2015structure}
J.~B. Swadling, D.~W. Wright, J.~L. Suter, P.~V. Coveney, Structure, dynamics,
  and function of the hammerhead ribozyme in bulk water and at a clay mineral
  surface from replica exchange molecular dynamics, Langmuir 31~(8) (2015)
  2493--2501.

\bibitem{thaplyal2015inverse}
P.~Thaplyal, A.~Ganguly, S.~Hammes-Schiffer, P.~C. Bevilacqua, Inverse thio
  effects in the hepatitis delta virus ribozyme reveal that the reaction
  pathway is controlled by metal ion charge density, Biochemistry 54~(12)
  (2015) 2160--2175.

\bibitem{gaines2016ribozyme}
C.~S. Gaines, D.~M. York, Ribozyme catalysis with a twist: {A}ctive state of
  the twister ribozyme in solution predicted from molecular simulation, J. Am.
  Chem. Soc. 138~(9) (2016) 3058--3065.

\bibitem{zhang2016assessing}
S.~Zhang, D.~R. Stevens, P.~Goyal, J.~L. Bingaman, P.~C. Bevilacqua,
  S.~Hammes-Schiffer, Assessing the potential effects of active site
  {Mg}$^{2+}$ ions in the {\it {glms}} ribozyme--cofactor complex, J. Phys.
  Chem. Lett. 7~(19) (2016) 3984--3988.

\bibitem{casalino2016activates}
L.~Casalino, G.~Palermo, U.~Rothlisberger, A.~Magistrato, Who activates the
  nucleophile in ribozyme catalysis? {A}n answer from the splicing mechanism of
  group {II} introns, J. Am. Chem. Soc. 138~(33) (2016) 10374--10377, {
  \\$^{\bullet{}\bullet{}}$ {QM/MM} free-energy calculations are here used to
  resolve the mechanism of phoshodiester hydrolysis in a large ribozyme. The
  authors explore a novel mechanism, where water molecules not only from the
  first coordination shell of {Mg}$^{2+}$ but also from the bulk participate in
  the cleavage reaction. Expensive CPMD simulations, implementing a fully
  Hamiltonian coupling scheme for QM/MM, are used}.

\bibitem{mlynsky2016inline}
V.~Ml{\'y}nsk{\'y}, G.~Bussi, Understanding in-line probing experiments by
  modeling cleavage of nonreactive {RNA} nucleotides, RNA 23~(5) (2017)
  712--720.

\bibitem{bingaman2017glcn6p}
J.~L. Bingaman, S.~Zhang, D.~R. Stevens, N.~H. Yennawar, S.~Hammes-Schiffer,
  P.~C. Bevilacqua, The {G}lc{N6P} cofactor plays multiple catalytic roles in
  the {\it {glms}} ribozyme, Nat. Chem. Biol. 13~(4) (2017) 439--445.

\bibitem{schuabb2017pressure}
C.~Schuabb, N.~Kumar, S.~Pataraia, D.~Marx, R.~Winter, Pressure modulates the
  self-cleavage step of the hairpin ribozyme., Nat. Commun. 8 (2017) 14661.

\bibitem{chen2017divalent}
H.~Chen, T.~J. Giese, B.~L. Golden, D.~M. York, Divalent metal ion activation
  of a guanine general base in the hammerhead ribozyme: {I}nsights from
  molecular simulations, Biochemistry 56~(24) (2017) 2985--2994.

\end{thebibliography}
